\documentclass[11pt,preprint]{aastex}

\begin{document}
\title{A dispersion-driven method for grant and proposal allocation}

\author{N. Mirabal and J. L. Contreras\\
Dpto. de F\'isica At\'omica, Molecular y Nuclear\\
Universidad Complutense de Madrid\\
Madrid, Spain}

\bigskip

{\bf Inspired by a recent editorial \citep{langer}, we
suggest a relative simple scheme to grade grants and proposals that might
enhance scientific innovation.}

Innovation should be at the forefront of research
practice. Unfortunately, it is easy to fall into
the habit of repeating the same type of experiment over and
time again \citep{langer}. The latter is not inherently bad since 
it guarantees 
safe returns, but it does not necessarily lead to innovation. 
Unfortunately, under most circumstances,  peer reviewers tend to lean heavily 
towards
ongoing experiments and improvements of verified results.
Recent analysis shows  that papers initially rejected 
received significantly more citations than those accepted on their 
first attempt 
\citep{calcagno}. These results highlight 
the valuable 
input received from reviewers, but it also reflects the dangerous side 
of scientific consensus.
Possible alternatives to the current funding scheme 
include choosing fundable proposals by
a lottery \citep{langer}, or providing funds to individual centers of
excellence without ties to specific goals \citep{loeb2}. 

We recognize that peer review is a necessary overseer and it is not
obvious that giving away
with it completely will provide direct improvement. However, some 
additional fine tuning
might help recognize innovation. New ideas tend to be controversial
and might take time to digest. In 1933 Fritz Zwicky had collected 
evidence to postulate dark matter \citep{zwicky}, but his work was 
not broadly accepted
until the 1970's.  Practically -- in many  modern day peer review rooms -- 
funding requests 
to carry out Zwicky's original 
research might have lead to a broad range of grades. 
It is reasonable to assume that at least one peer
reviewer might have deemed Zwicky's findings extraordinary.  
But low grades might have blurred the high grade and probably Zwicky's proposal
would not have passed the first
round of discussions. A direct way to reveal this type of anomaly is to
measure the dispersion of the grades, in its simplest version through  
their standard deviation. Assuming the qualifications of all reviewers
are equal, 
we suggest that the final decision on grants and proposals should not
only consider average grade but also its dispersion.

But how to allocate funds in this fashion? For that, 
\citet{loeb} has advanced a reasonable breakdown into safe and
risky investments that might guide
us here. Adopting such division, 80\% of grant/time allocation would
go to  secure research {\it i.e.} proposals with the highest average grades. 
The remaining 20\% could be divided among the risky but with potential
high gains {\it i.e.} proposals with the
highest standard deviations.  Percentages and weights given to each
variable can be adjusted if needed. 

\clearpage

Certainly, a number of risky ideas that turn out to be groundbreaking in
the long run might go unnoticed if no reviewer deems it extraordinary. 
For those, only perseverance and resubmission will find a reward.
Finally, an argument could be made that such scheme would 
improve the health of the reviewing system.
Presently, reviewers understand the devastating effect that a single bad
grade has on the fate of a proposal, and may feel pressed to 
artificially
inflate their grades. 
Knowing that an alternative scheme is in place 
might relieve them from this pressure and make 
grading more reliable. On the other side, 
proposers might also have greater incentive to be creative 
if there are improved odds for innovative proposals.
 
This scheme is by no means the only solution to the problem nor does 
it guarantee
innovation. At present, it is just a modest proposal
to enrich the pool of ideas meant to help promote it. 
\bigskip

\acknowledgements 
N.M. acknowledges support from the Spanish government
through a Ram\'on y Cajal fellowship.

\end{document}